\documentclass[aps,prb,groupedaddress, twocolumn, notitlepage, portrait]{revtex4-2}
\usepackage[utf8]{inputenc}
\usepackage{amsmath}
\usepackage{amsfonts}
\usepackage{amssymb}
\usepackage{braket}
\usepackage{graphicx}
\usepackage[colorlinks=true, citecolor=blue, linkcolor=blue, urlcolor=blue]{hyperref}

\begin{document}
\begin{abstract}
We search for short-range Hamiltonians of finite spin-1 kagome systems, maximizing the overlaps with lattice Moore-Read states. Our starting point is an exact, long-range parent Hamiltonian for such a state on a finite plane, obtained from conformal field theory. A truncation procedure is applied to it, which retains only short-range terms and makes it easy to define the Hamiltonian on a torus. Finally, the remaining coefficients are optimized, to yield maximized overlaps between exact diagonalization results and model ground states. In the best cases, the squared overlaps exceed 0.9 and 0.8 for the three lowest states of 12- and 18-site systems, respectively, suggesting that the obtained Hamiltonians are good parent Hamiltonians for a non-Abelian topological order.
\end{abstract}

\title{Non-Abelian chiral spin liquid on spin-1 kagome lattice: truncation of an exact Hamiltonian and numerical optimization}

\author{B\l a\.{z}ej Jaworowski}
\affiliation{Department of Physics and Astronomy, Aarhus University, DK-8000 Aarhus C, Denmark}
\affiliation{Max-Planck-Institut f\"{u}r Physik komplexer Systeme, D-01187 Dresden, Germany}
\email{blazej@phys.au.dk}
\author{Anne E. B. Nielsen}
\affiliation{Department of Physics and Astronomy, Aarhus University, DK-8000 Aarhus C, Denmark}
\affiliation{Max-Planck-Institut f\"{u}r Physik komplexer Systeme, D-01187 Dresden, Germany}

\maketitle

\section{Introduction}

Two-dimensional topological orders \cite{wen1990topological} are expected to have anyonic excitations \cite{leinaas1977theory,wilczek1982quantum,nayak2008nonabelian}. 
Recent experiments \cite{bartolomei2020fractional,nakamura2020direct} support the theoretical prediction for the existence of Abelian anyons as excitations above the fractional quantum Hall states in a two-dimensional electron gas (2DEG) in high magnetic fields. 
The presence of the non-Abelian anyons, which may provide a platform for fault-tolerant quantum computation \cite{kitaev2003fault,sarma2005topologically, nayak2008nonabelian}, is more challenging to demonstrate experimentally, yet indirect measurements suggest that they indeed exist \cite{tiemann2012unraveling,banerjee2018observation}.

The quantum Hall states in a 2DEG are usually described in terms of a continuum formalism \cite{laughlin}.  However, theoretical works show that the states with the same topological orders can also be constructed in lattice models
\cite{kalmeyer1987equivalence,moller2009composite,kapit2010exact,Tang,SunNature,Neupert,nielsen2013local}, which can be seen as spin systems, and their ground states as chiral quantum spin liquids. It is expected that they can be realized in quantum simulators, e.\ g. optical lattices \cite{sorensen2005fractional,palmer2006high,hafezi2007fractional,palmer2008optical,
moller2009composite,kapit2010exact,you2010quantum,schmied2011quantum,yao2013realizing,
cooper2013reaching,nielsen2013local}. The hopes that this is the case are raised by recent experiments on spin liquids characterized by a different, nonchiral type of topological order. One such model was recently engineered in an array of Rydberg atoms \cite{semeghini2021probing}. Also, a nonchiral topological quantum spin liquid state was simulated in a quantum computer \cite{lu2009demonstrating,song2018demonstration,satzinger2021realizing}.

In quantum simulators, one has much greater control over the system parameters than in solid state systems. The possibility of addressing individual sites in such a setting allows for novel ways of manipulating anyons. While a majority of proposed and realized experiments on quantum Hall anyons in 2DEG focus on the anyons within the edge states and/or pinned by static impurities in the bulk \cite{kivelson1990semiclassical,martin2004localization, sarma2005topologically,venkatachalam2011local,papic2018imaging,bartolomei2020fractional,nakamura2020direct, feldman2021fractional, carrega2021anyons}, the proposals in which one can move the bulk anyons in a controlled way (e.g. to braid them) are rare \cite{grass2018optical}. For quantum simulators, there are more proposals of the latter type \cite{zhang2007anyonic, han2007scheme,aguado2008creation, jiang2008anyonic, kapit2012nonabelian}. Moreover, such experiments were actually performed in quantum computers \cite{lu2009demonstrating,song2018demonstration,satzinger2021realizing}. Given the perspectives for realization and manipulation of anyons in quantum simulators, it is desirable to invent further lattice models with different types of topologically ordered ground states.

Let us focus on the chiral spin liquid with the same topological order as exhibited by the bosonic Moore-Read quantum Hall state with $\nu=1$ filling factor. In addition to providing a platform for non-Abelian anyons, a realization of such a phase would be interesting, because bosonic quatum Hall states by definition cannot exist in a 2DEG, which is a system of fermions. Long-range spin-1 Hamiltonians on two-dimensional lattices were proposed, for which the ground state is exactly described by a discretized Moore-Read wavefunction \cite{nielsen2011quantum,greiter2014parent,glasser2015exact}. 

While these models are of significant theoretical importance, the experimentally relevant interactions typically are short-range. The bosonic Moore-Read topological order is predicted to arise in some topological flat-band models with short-range hoppings \cite{wang2012nonabelian, cooper2013reaching,sterdyniak2012particle}. A short-range Hamiltonian generating a spin liquid with this kind of topological order was also constructed on a spin-1 triangular lattice \cite{greiter2009nonabelian}. A spin-1 square lattice model, based on a truncation of a long-range Hamiltonian, was proposed in Ref.\ \cite{glasser2015exact}, and studied further in \cite{chen2018nonabelian}.  Moreover, a general procedure of truncating the long-range terms was formulated for Hamiltonians constructed from conformal field theory \cite{nandy2019truncation}. While so far it was applied only to the parent Hamiltonians of Abelian states, the long-range Hamiltonian from Ref.\ \cite{glasser2015exact} has a similar structure, which makes the application of this method possible.

Here, we instead study the kagome lattice. The kagome lattice plays an important role in the search for quantum spin liquids, due to its geometric frustration. Various kinds of spin liquids were studied in models defined on this lattice \cite{lecheminant1997order, yan2011spin, depenbrock2012nature, iqbal2013gapless,he2014chiral,gong2014emergent,bauer2014chiral,gong2015global}. It is thus natural to ask if the non-Abelian Moore-Read topological order can also be realized on the kagome lattice. A study of a certain spin-1 kagome model provides some hints that this could be the case \cite{liu2018nonabelian}.

In this work, we construct a local Hamiltonian on the spin-1 kagome lattice and study it using exact diagonalization. It is created by combining the truncation \cite{nandy2019truncation} of an exact model based on conformal field theory (CFT) \cite{glasser2015exact} with numerical optimization based on exact diagonalization.  We show that its three lowest-energy states have significant overlaps with model Moore-Read-like states in intermediate-size systems. 

We choose the overlap as an indicator of the topology, because: (i) it provides a more convincing signature of topological order than the energy spectrum alone -- the energy spacings resembling a topological quasi-degeneracy in a single system may occur by accident, and the limitations of the exact diagonalization do not allow to compute the scaling of the energy gap with system size, (ii) the overlap is relatively easy to compute (it requires only a single diagonalization, in contrast to the spectral flow or the many-body Chern number, for which one performs many diagonalizations, each for a different value of the boundary phase) (iii) it is a single, non-quantized number, which can be easily fed into the optimization procedure, and (iv) there are situations where it is the excited states, instead of the ground states, which have high overlaps with the model states -- this suggests that the system is close to a topological phase, even though other signatures might not be visible (see e.g. \cite{nandy2019truncation}). High overlaps suggest that the states are topological, but the reverse is not the case. The system can possess the desired topological order even if the overlaps are low. Such regions of the parameter space will remain undetected within our approach.

The paper is organized as follows. In Sec.\ \ref{sec:wfn}, we define our system and recall the lattice Moore-Read wavefunctions on a torus \cite{zhang2021resonating}, serving as a reference in our computations. Next, in Sec.\ \ref{sec:truncation}, we review the truncation procedure proposed in \cite{nandy2019truncation} and apply it to the exact parent Hamiltonian of the lattice Moore-Read state. Then, in Sec.\ \ref{sec:optimization}, we optimize the coefficients of the truncated Hamiltonian, obtaining states with large overlaps with model wavefunctions, and a visible topological degeneracy. Sec.\ \ref{sec:discussion} discusses the results and their relations to other works. Section \ref{sec:conclusions} concludes the article.

\begin{figure}
\includegraphics[width=0.45\textwidth]{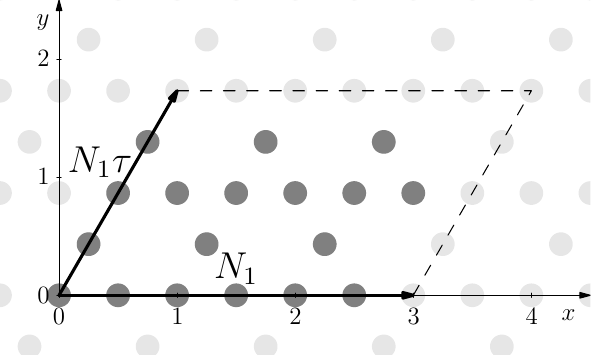}
\caption{The site positions in our notation, on the example of an $N_1 \times N_2 = 3 \times 2$ system. The gray dots denote the sites on an infinite plane, with complex positions $z_j=x_j+iy_j$. A parallelogram (the arrows and dashed lines) containing $N_1\times N_2$ unit cells ($3 N_1N_2$ sites -- the dark gray ones) is chosen and transformed into a torus. On the complex plane, the torus is spanned by the vectors $N_1$ and $N_1\tau$ (arrows), with $|N_1\tau|=N_2$.
}
\label{fig:lattice}
\end{figure}

\section{The system and the model states}\label{sec:wfn}

In the continuum, one can construct a series of non-Abelian Moore-Read states from correlation functions of conformal fields \cite{moore1991nonabelions}. The bosonic $\nu=1$ case can be modified to describe a non-Abelian chiral spin liquid in a lattice of spin-1 sites on a plane \cite{greiter2009nonabelian,glasser2015exact} or on a torus \cite{greiter2014parent,zhang2021resonating}. In this Section, we recall the expressions for the lattice states from \cite{glasser2015exact,zhang2021resonating}.

Let us first define the system we work on and the notation we use to describe it. We start from an infinite kagome lattice on a plane. We introduce a complex position for each site $j$: $z_j=x_j+iy_j$, where $x_j, y_j$ are the Cartesian coordinates of the site. We fix the lattice constant to unity. Then, we choose a parallelogram of size $N_1\times N_2$ in the direction of the two kagome lattice vectors $\mathbf{a}_1=[1,0]$, $\mathbf{a}_2=[\frac{1}{2},\frac{\sqrt{3}}{2}]$ (see Fig.\ \ref{fig:lattice}). The vectors spanning the parallelogram can be written as complex numbers $N_1$, $N_1\tau$, where $\tau=\frac{N_2}{N_1}(\frac{1}{2}+i\frac{\sqrt{3}}{2})$.  We number the sites within it by $j=1, 2,\dots, N$, with $N=3N_1N_2$. On each of them, we introduce a spin-1, with three possible values of the $S_z$ component: $s_j\in\{-1,0,1\}$. The spin configurations will be denoted as $\ket{\mathbf{s}}=\ket{s_1s_2\dots s_N}$.

Before discussing the case of the torus, let us recall the lattice Moore-Read wavefunction for open boundary conditions (plane), given by \cite{glasser2015exact}
\begin{equation}
\ket{\psi_{\mathrm{OBC}}}=\frac{1}{C}\sum_{\mathbf{s}}\psi_{\mathrm{OBC}}(\mathbf{s})\ket{\mathbf{s}}, 
\end{equation}
where $C$ is the normalization constant, and the coefficients $\psi_{\mathrm{OBC}}(\mathbf{s})$ are given by
\begin{multline}
\psi_{\mathrm{OBC}}\left(\mathbf{s}\right)
=\delta\left(\sum_js_j\right)(-1)^{\sum_{k=1}^{N/2}s_{2k-1}}\times \\ \times \mathrm{Pf}_0\left(\frac{1}{z_j-z_k}\right)\prod_{j<k}\left(z_j-z_k\right)^{s_js_k}
\label{eq:modelstate_plane}
\end{multline}
Here, $\delta$ is the Kronecker delta which enforces $\sum_js_j=0$, and $\mathrm{Pf}_0$ means that the Pfaffian includes only the coordinates $z_j$, $z_k$ of the sites with $s_j=s_k=0$.

In small kagome systems, open boundary conditions make it difficult to extract bulk properties, because a majority of the sites lies on the edges (in contrast to e.g. the square lattice studied in \cite{glasser2015exact}). Thus, we prefer a geometry without edges, such as a torus. The toroidal geometry is also convenient for the exact-diagonalization calculations, because it allows us to use the conservation of the lattice momentum to decrease the numerical complexity of the problem. Thus, in the following, we transform our parallelogram into a torus by gluing its opposite edges, and recall the generalizaton of \eqref{eq:modelstate_plane} to toroidal systems, constructed in Ref.\ \cite{zhang2021resonating}.
 
It is helpful to introduce rescaled coordinates on the torus, $\xi_j=z_j/N_1$. In these coordinates, the vectors spanning the torus are $1$, $\tau$. To express the model wavefunctions from \cite{zhang2021resonating}, we need the Jacobi theta function with characteristics,
\begin{multline}
\theta
\begin{bmatrix}
a \\ b
\end{bmatrix}
\left(\xi|\tau\right)=\\
=\sum_{n\in \mathbb{Z}} \exp\left(i\pi\tau\left(n+a\right)^2+2\pi i\left(n+a\right)\left(\xi+b\right)\right),
\label{eq:theta_char}
\end{multline}
where $a$ and $b$ are real parameters. In terms of \eqref{eq:theta_char}, we define
\begin{align}
&\theta_1\left(\xi|\tau\right)=\theta
\begin{bmatrix}
1/2 \\ 1/2
\end{bmatrix}
\left(\xi|\tau\right),&
\theta_2\left(\xi|\tau\right)=\theta
\begin{bmatrix}
1/2 \\ 0
\end{bmatrix}
\left(\xi|\tau\right), \nonumber \\
&\theta_3\left(\xi|\tau\right)=\theta
\begin{bmatrix}
0 \\ 0
\end{bmatrix}
\left(\xi|\tau\right),&
\theta_4\left(\xi|\tau\right)=\theta
\begin{bmatrix}
0 \\ 1/2
\end{bmatrix}
\left(\xi|\tau\right).
\end{align}
Also, we define the functions
\begin{equation}
E(\xi_j-\xi_k|\tau)=\frac{\theta_{1}(\xi_j-\xi_k|\tau)}{\partial_\xi\theta_1(\xi|\tau)|_{\xi=0}},
\label{eq:Edef}
\end{equation}
\begin{equation}
\mathcal{P}_\mu(\xi_j-\xi_k| \tau)=\frac{\theta_{\mu+1}(\xi_j-\xi_k|\tau)\partial_\xi\theta_1(\xi|\tau)|_{\xi=0}}{\theta_{\mu+1}(0|\tau)\theta_1(\xi_j-\xi_k|\tau)},
\label{eq:Pdef}
\end{equation}
where $\mu\in\{1,2,3\}$. 
The three model wavefunctions on the torus are given by,
\begin{equation}
\ket{\psi_\mu}=\frac{1}{C_\mu}\sum_{\mathbf{s}}\psi_\mu(\mathbf{s})\ket{\mathbf{s}}, 
\end{equation}
with $\mu=1,2,3$, $C_\mu$ being the normalization constant, and the unnormalized coefficient $\psi_\mu(\mathbf{s})$ given by \cite{zhang2021resonating}
\begin{multline}
\psi_\mu\left(\mathbf{s}\right)=\\
=\delta\left(\sum_js_j\right) 
(-1)^{\sum_{k=1}^{N/2}s_{2k-1}}
\mathrm{Pf}_0\left(\mathcal{P}_\mu\left(\xi_j-\xi_k| \tau\right) \right)\times \\ 
\times\frac{\theta_{\mu+1}(\sum_{j}\xi_js_j|\tau)}{\theta_{\mu+1}(0|\tau) }\prod_{j<k}E\left(\xi_j-\xi_k|\tau\right)^{s_js_k}.
\label{eq:modelstate}
\end{multline}
Here, again, $\delta$ is the Kronecker delta enforcing $\sum_js_j=0$, and $\mathrm{Pf}_0$ means that the Pfaffian includes only the coordinates $\xi_j$, $\xi_k$ of the sites with $s_j=s_k=0$.

The three wavefunctions \eqref{eq:modelstate} explicitly realize the threefold topological degeneracy of the Moore-Read-type topological order. That these states have the same topological order as the continuum states from \cite{moore1991nonabelions} was confirmed in \cite{zhang2021resonating} by evaluating the modular $S$ and $T$ matrices.

In general, there is some phase freedom in the choice of the wavefunctions: multiplying them by certain $\mathbf{s}$-dependent phase factors does not change their topological order. Thus, if one wants to compute the overlap with eigenstates obtained from exact diagonalization, one has to fix the phases correctly. In Eq.\ \eqref{eq:modelstate}, the phases are fixed such that the state is a singlet. As long as the Hamiltonian in the exact diagonalization is $SU(2)$-symmetric, we can use Eq.\ \eqref{eq:modelstate} -- which is the case in the model that we define in Sec.\ \ref{sec:truncation}.

We finally compare the wavefunctions introduced in \cite{greiter2014parent} and \cite{zhang2021resonating}. For kagome systems of sizes $1\times 2$ and $2\times 2$, we verified numerically that the wavefunctions in \eqref{eq:modelstate} and the torus wave functions in \cite{greiter2014parent} span the same subspace of the Hilbert space, if an appropriate gauge is chosen. We also note that while the overlaps are perfect on the torus, this is not true on a finite plane -- i.e. \eqref{eq:modelstate_plane} is not equivalent to the planar wavefunction in \cite{greiter2009nonabelian,greiter2014parent} for finite $N$. This is due to a different choice of the placement of the background charge. As $N\rightarrow \infty$, the two planar wavefunctions approach each other \cite{glasser2015exact}.

\section{Results: truncated CFT Hamiltonian}\label{sec:truncation}

We now look for an approximate parent Hamiltonian for the states \eqref{eq:modelstate} on the kagome lattice. Let us again start by looking at a planar system (i.e. with open boundary conditions). Refs.\ \cite{nielsen2011quantum,greiter2014parent,glasser2015exact} proposed the following planar Hamiltonian,
\begin{equation}
H=\sum_{a\in \{x,y,z\}}\sum_{j=1}^{N}\Lambda_j^{a\dagger}\Lambda_j^{a},
\end{equation}
where $\Lambda_j^{a}$ is an operator annihilating the lattice Moore-Read state, 
\begin{multline}
\Lambda_j^{a}=\sum_{k(\neq j)}\omega_{jk}\left[
\frac{2}{3}S^{a}_k-\frac{5}{12}i\left(\sum_{b,c}\epsilon_{abc}S^{b}_kS^{c}_j\right)\right.-\\
-\left.\frac{1}{12}\left(\sum_b\left(S_j^aS_j^b+S_j^bS_j^a\right)
S_k^b\right) \right].
\label{eq:lambda}
\end{multline}
Here, $k(\neq j)$ means that we sum over all possible values of $k$ except from $j$, $\epsilon_{abc}$ is the Levi-Civita tensor, $\omega_{jk}=\frac{1}{z_j-z_k}$, and $S_j^{a}$, $a\in\{x,y,z\}$, is the operator of the $a$th component of the spin, acting on site $j$.

The resulting Hamiltonian is (up to an additive constant)
\begin{multline}
H = \frac{1}{3}\sum _{j\ne k}\sum_a\left({\omega }_{{jk}}^{*}{\omega }_{{jk}}+2 \sum _{l(\ne j,k)}{\omega }_{{lj}}^{*}{\omega }_{{lk}}\right){S}_{j}^{a}{S}_{k}^{a} -\\
- \frac{1}{6} \sum _{j\ne k}{\omega }_{{jk}}^{*}{\omega }_{{jk}}{\left(\sum_a{S}_{j}^{a}{S}_{k}^{a}\right)}^{2}+\\
+\sum _{j\ne k\ne l} \sum_{a,b}\left( \frac{1}{3}{\omega }_{{jl}}^{*}{\omega }_{{jk}}-\frac{1}{2}{\omega }_{{jl}}{\omega }_{{jk}}^{*}\right){S}_{j}^{a}{S}_{k}^{a}{S}_{j}^{b}{S}_{l}^{b}.
\label{eq:CFTham}
\end{multline}
Equation \eqref{eq:CFTham} was constructed in \cite{nielsen2011quantum,glasser2015exact} using conformal field theory, thus we call it a ``CFT Hamiltonian'' throughout this work. It was also shown in \cite{glasser2015exact} that the exact ground state of \eqref{eq:CFTham} is the planar wavefunction \eqref{eq:modelstate_plane}. We stress that, since \eqref{eq:CFTham} is not defined on a torus, it is not a parent Hamiltonian of the torus wavefunctions \eqref{eq:modelstate}.

It can be shown that \eqref{eq:CFTham} is equal to the planar Hamiltonian from Ref.\ \cite{greiter2014parent} (up to multiplicative and additive constants): the imaginary part of the last term of \eqref{eq:CFTham} corresponds to the triple-product term from \cite{greiter2014parent} and the real part corresponds to the other three-spin term from \cite{greiter2014parent}. We note that the planar Hamiltonian in \cite{greiter2014parent} was proposed as an approximate parent Hamiltonian of a planar wavefunction different from \eqref{eq:modelstate_plane}, becoming exact for an infinite system. In \cite{glasser2015exact}, it was shown that the true ground state of \eqref{eq:CFTham} is \eqref{eq:modelstate_plane}. These two results are consistent, as the planar wavefunction from \cite{greiter2014parent} approaches \eqref{eq:modelstate_plane} as $N\rightarrow \infty$.

The Hamiltonian \eqref{eq:CFTham} has long-range terms, with coefficients depending on the system size and shape. Thus, it is unclear how one should wrap it around the torus if periodic boundary conditions were applied. To generate a local Hamiltonian depending only on the local structure of the lattice, we proceed by following the truncation procedure, devised in Ref.\ \cite{nandy2019truncation} for parent Hamiltonians of lattice Laughlin states.
The idea is that, while in \eqref{eq:CFTham} the coefficients of the two-spin terms depend on the positions of all sites in the system, in \eqref{eq:lambda} all the coefficients depend only on the vector $z_j-z_k$ connecting the sites $j$ and $k$ on which the given term acts. Thus, if we want to have a result that depends only on the local environments of the involved sites, the truncation has to be performed on the level of $\Lambda_j^{a}$.

The truncation is performed simply by substituting
\begin{equation}
 \omega_{jk}\rightarrow\tilde{\omega}_{jk}=\begin{cases}
			\omega_{jk} & \text{if $|z_j-z_k|\leq r_{\mathrm{trunc}}$}\\
			0 & \text{if $|z_j-z_k|> r_{\mathrm{trunc}}$}\\
		 \end{cases}
\label{eq:omega}
\end{equation}
to \eqref{eq:lambda}, and in consequence to \eqref{eq:CFTham}. The Hamiltonian after truncation remains $SU(2)$-symmetric.

\begin{figure}
\includegraphics[width=0.5\textwidth]{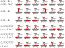}
\caption{The terms and parameters of the truncated Hamiltonian $H_{\mathrm{tr}}(\mathbf{J})$. The red dots and lines denote the involved sites and bonds between them, respectively. In the case of three-spin terms, the arrows go $k\rightarrow j \rightarrow l$ (this is important when considering complex parameters), with $+ k\leftrightarrow l$ meaning that we have to consider also terms of the same form and $k$ exchanged with $l$ (and $+\mathrm{c.c.}, k\leftrightarrow l$ means that these terms should be complex conjugated). Summation over $a,b$ indices is implied. The dashed gray parallelograms are the boundaries of the kagome unit cell.
}
\label{fig:elements}
\end{figure}

In \eqref{eq:omega}, $r_{\mathrm{trunc}}$ is a truncation radius. In this work we consider $r_{\mathrm{trunc}}=1/2$, which is the smallest possible value available in the kagome lattice. If the system is big enough, the remaining terms in the bulk can be parametrized as in Fig.\ \ref{fig:elements} (here, $J_{\mathrm{t1}}$ and $J_{\mathrm{t2}}$ are complex, while all the other coefficients are real). We set the energy scale (i.e. multiply the Hamiltonian by a constant) by fixing $J_{\mathrm{h1}}=1$, and denote the set of remaining parameters as $\mathbf{J}=[J_{\mathrm{s1}},J_{\mathrm{h2}},J_{\mathrm{h3}},J_{\mathrm{t1}},J_{\mathrm{t2}},J_{\mathrm{t3}}]$. The Hamiltonian described by Fig.\ \ref{fig:elements} is then named $H_{\mathrm{tr}}(\mathbf{J})$. The values of $\mathbf{J}$ resulting from the truncation procedure are denoted by $\mathbf{J}_0$ and provided in Table \ref{tab:elementsCFT}. In contrast to the non-truncated Hamiltonian, $H_{\mathrm{tr}}(\mathbf{J}_0)$ (and $H_{\mathrm{tr}}(\mathbf{J})$ in general) can be easily defined on a torus simply by repeating these terms periodically and applying periodic boundary conditions.

The ground state of the truncated Hamiltonian on the torus (as well as on the plane, but we focus on the torus here) is not known. Thus, we diagonalize $H_{\mathrm{tr}}(\mathbf{J}_0)$ numerically for toroidal systems of sizes $1\times 2$, $2\times 2$ and $3\times 2$ unit cells. For the diagonalization, we use the Lanczos method implemented in Fortran using the ARPACK package, with the matrix-vector multiplication parallelized using OpenMP. We use the $S_z$ conservation and diagonalize only the $\sum_j s_j=0$ block of the Hamiltonian.

To check if $H_{\mathrm{tr}}(\mathbf{J}_0)$ generates the Moore-Read states, we calculate the squared overlaps of its low-energy eigenstates with the model states \eqref{eq:modelstate}. Because of the threefold topological degeneracy of the model states, we define the squared overlap for an exact eigenstate $\ket{\phi}$ of the truncated Hamiltonian as
\begin{equation}
O(\ket{\phi})=\left|\braket{\phi|\tilde{\psi}_1}\right|^2+\left|\braket{\phi|\tilde{\psi}_2}\right|^2+\left|\braket{\phi|\tilde{\psi}_3}\right|^2,
\label{eq:overlap}
\end{equation}
where $\ket{\tilde{\psi}_\mu}$ are orthonormalized versions of the $\ket{\psi_\mu}$. The orthogonalization is necessary because for finite systems the $\ket{\psi_\mu}$ states are not always orthogonal. We have $|\braket{\psi_\mu|\psi_\nu}|<0.002$, $|\braket{\psi_\mu|\psi_\nu}|\approx 0.075$ and $|\braket{\psi_\mu|\psi_\nu}|<0.115$ for $1\times 2$, $2\times 2$ and $3\times 2$ systems, respectively, where $\mu\neq \nu$.

\begin{figure}
\includegraphics[width=0.45\textwidth]{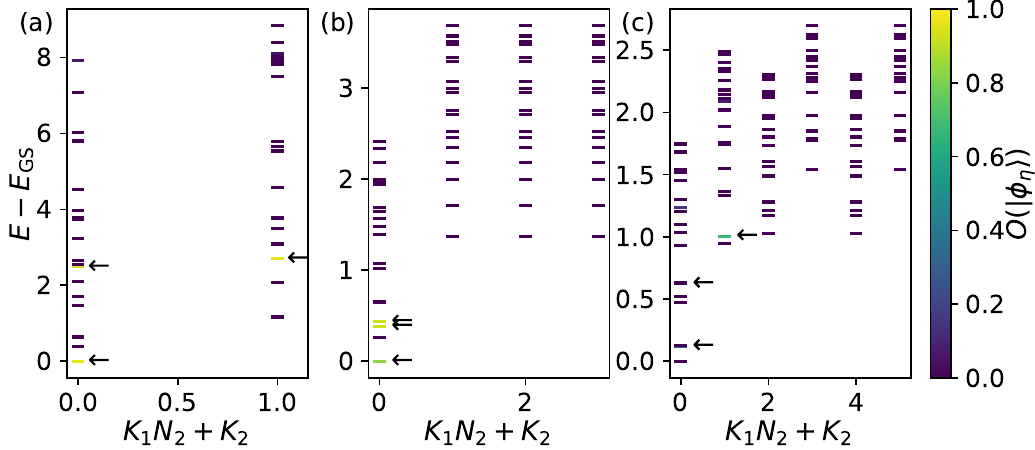}
\caption{Momentum-resolved low-energy spectra of the truncated CFT Hamiltonian $H_{\mathrm{tr}}(\mathbf{J}_0)$. Subfigiures (a), (b), (c) show the results for system sizes $1\times 2$, $2\times 2$, $3\times 2$, respectively. The energy is measured with respect to the ground state energy $E_{\mathrm{GS}}$. $K_1$ and $K_2$ are the total integer lattice momenta in the directions of lattice vectors $\mathbf{a}_{1}$, $\mathbf{a}_{2}$, respectively, with $K_1\in \{0,1,\dots, N_1-1\}$ and $K_2\in \{0,1,\dots, N_2-1\}$. The color denotes the squared overlap $O(\ket{\phi_\eta})$ \eqref{eq:overlap} for the given state $\ket{\phi_\eta}$. Note that in (c) the two states with highest overlaps for $\mathbf{K}=[0,0]$ are close to other states with almost zero overlaps, so they are not easily visible. Thus, we point out the three states with highest overlaps in each plot using black arrows.}
\label{fig:cftspectra}
\end{figure}

In Table \ref{tab:elementsCFT}, we show the squared overlaps \eqref{eq:overlap} for the three investigated systems. We computed 20 lowest-energy states per momentum subspace. Table \ref{tab:elementsCFT} contains the three maximal squared overlaps from all the computed states of a given system. The three states with largest overlaps are not necessarily the ground states, as can be seen in the energy spectra plotted in Fig.\ \ref{fig:cftspectra}. Moreover, for the $3\times 2$ system, one of the squared overlaps is actually very small -- about 0.3. Thus, our results do not indicate clearly the presence of the spin liquid phase for $H_{\mathrm{tr}}(\mathbf{J}_0)$. However, we also do not rule it out, because an energy gap and a topological degeneracy might emerge for larger systems, and, as mentioned in the Introduction, a state can still be in a given topological class even though the overlap is small. 

\begin{table}[htbp]
(a) Parameters
\begin{ruledtabular}
\begin{tabular}{llllllll}
 \multicolumn{1}{l}{$J_{\mathrm{s1}}$} & \multicolumn{1}{l}{$J_{\mathrm{h2}}$} & \multicolumn{1}{l}{$J_{\mathrm{h3}}$} & \multicolumn{1}{l}{$\mathrm{Re}J_{\mathrm{t1}}$} & \multicolumn{1}{l}{$\mathrm{Im}J_{\mathrm{t1}}$} & \multicolumn{1}{l}{$\mathrm{Re}J_{\mathrm{t2}}$} & \multicolumn{1}{l}{$\mathrm{Im}J_{\mathrm{t2}}$} & \multicolumn{1}{l}{$J_{\mathrm{t3}}$} \\ \hline 
$-\frac{1}{4}$ & $-\frac{1}{2}$ & $-1$ & $-\frac{1}{16}$ & $\frac{5}{16}\sqrt{3}$ & $\frac{1}{16}$ & $\frac{5}{16}\sqrt{3}$ & $\frac{1}{8}$ \\
\end{tabular}
\end{ruledtabular}
~\\~\\
(b) Squared overlaps
\begin{ruledtabular}
\begin{tabular}{lllllllll}
\multicolumn{3}{l}{$1\times 2$}& \multicolumn{3}{l}{$2\times 2$}& \multicolumn{3}{l}{$3\times 2$}  \\ \hline
0.964 & 0.960 & 0.939 & 0.926 & 0.925 & 0.834 & 0.821 & 0.669 & 0.332 \\ 
\end{tabular}
\end{ruledtabular}
\caption{(a) The values of the parameters $\mathbf{J}_0$ of the truncated CFT Hamiltonian $H_{\mathrm{tr}}(\mathbf{J}_0)$ (see Fig.\ \ref{fig:elements}). (b) The three maximum squared overlaps \eqref{eq:overlap} from the 20 lowest-energy states per momentum subspace, sorted in the decreasing order, for the three investigated system sizes, with Hamiltonian parameters shown in (a).}
\label{tab:elementsCFT}
\end{table}

\section{Results: overlap optimization}\label{sec:optimization}
It may be tempting to increase the truncation radius in hope for improving the results. However, such a Hamiltonian would have further-range terms, which would make it less convenient to work with. Also, such terms would not necessarily generate better results in finite systems with periodic boundary conditions, where they would wrap around the torus, possibly many times. Thus, we use a different approach: we continue with the Hamiltonian $H_{\mathrm{tr}}(\mathbf{J})$, but allow the coefficients $\mathbf{J}$ to vary. We optimize them to maximize the squared overlap \eqref{eq:overlap} with the model states.

Because we expect three topologically ordered ground states, there are many ways to define the cost function to minimize. We choose the following one. Let us define states $\ket{\phi_\eta (\mathbf{J})}$ as the exact diagonalization eigenstates, sorted according to the squared overlap $O(\ket{\phi_\eta(\mathbf{J})})$ in a decreasing order (note that in the exact diagonalization, we do not obtain the full spectrum, but only a few lowest eigenstates). Then, we define the following cost function 
\begin{equation}
F\left(\mathbf{J}\right)=3-O\left(\ket{\phi_1(\mathbf{J})}\right)-O\left(\ket{\phi_2(\mathbf{J})}\right)-O\left(\ket{\phi_3(\mathbf{J})}\right).
\label{eq:cost}
\end{equation}
In other words, if we define the projectors $\Pi_{\mathrm{ED}}(\mathbf{J})=\sum_{\eta=1}^{3}\ket{\phi_\eta(\mathbf{J})}\bra{\phi_\eta(\mathbf{J})}$ and $\Pi_{\mathrm{model}}=\sum_{\mu=1}^{3}\ket{\tilde{\psi}_\mu}\bra{\tilde{\psi}_\mu}$, $F$ can be expressed as
\begin{equation}
F\left(\mathbf{J}\right)=3-\mathrm{Tr}\left(\Pi_{\mathrm{ED}}\left(\mathbf{J}\right)\Pi_{\mathrm{model}}\right).
\end{equation}
It takes values between zero and three. We use the notation $\ket{\phi_\eta}$ (without the argument $\mathbf{J}$) for the final states $\ket{\phi_\eta (\mathbf{J})}$ at the optimal parameters.

We use the simplicial homology global optimization (SHGO) algorithm, implemented in the \verb|scipy.optimize.shgo| function of the SciPy library (version 1.6.2). The optimization is performed by a Python script calling the Fortran exact diagonalization code. The largest system size for which we can run the script is $2\times 2$, because for the $3\times 2$ system a single diagonalization takes about 17 hours on our high-performance computing cluster, and within a single optimization run it may be called several hundred times. Thus, we perform the optimization for the $2\times 2$ system. Since all the model states for this system size have momentum $\mathbf{K}=[0,0]$, during the optimization we diagonalize the Hamiltonian only within this subspace, and use its 20 lowest states to compute $F\left(\mathbf{J}\right)$. After the optimization, the calculations for the obtained optimal $\mathbf{J}$ are performed also for other subspaces and system sizes, in order to determine the presence of topological degeneracy and check if the results are consistent as a function of the system size. 

The Hamiltonian depends on eight real parameters. To simplify the optimization process and to test which parameters are important, we perform a number of optimization runs, in which different sets of parameters are set to zero. The best results are shown in Table \ref{tab:CFTopt} (a). Here, ``NA'' means that the given parameter is fixed at zero. The ``TD'' column shows if the systems exhibit the topological degeneracy ($\checkmark$) or not ($\times$), with the first, second and third symbol corresponding to $1\times 2$, $2\times 2$, $3\times 2$ system, respectively. For the purpose of this table, we say that a system exhibits the topological degeneracy if the three states $\ket{\phi_1}$, $\ket{\phi_2}$, $\ket{\phi_3}$ with highest squared overlaps $O(\ket{\phi_\eta})$ are the three lowest states. Note that this definition is much less strict than the proper definition of the topological degeneracy -- the energy split between these states can be arbitrarily large, the gap between the third and fourth state can be arbitrarily small, and we do not try to deduce the systems' behavior in the thermodynamic limit.

In some optimization runs, the resulting parameters were particularly ``round'' (e.g. 0 or 0.6), and lying precisely at the upper or lower bound or halfway between them. This suggests that the algorithm did not find a true local minimum (and in consequence, also the global one). In an attempt to avoid this effect, we studied two sets of bounds, denoted by ``A'' or ``B'' in the first column of Table \ref{tab:CFTopt} (a). The values of these bounds are shown in Table \ref{tab:CFTopt} (b). We performed all the optimization runs for both sets of bounds, and included only the best results in Table \ref{tab:CFTopt} (a). However, one can see that this does not always help. Thus, for every row of Table \ref{tab:CFTopt} (a), we performed an additional local optimization using the Nelder-Mead method (the \verb|scipy.optimize.minimize| function), taking the result of the SHGO procedure as a starting point. For the cases when SHGO returned one or more ``round'' parameter values, this indeed yielded significantly different results, which we show in Table \ref{tab:CFTopt} (c). For other ones, the local optimization confirmed that SHGO indeed found a local minimum.

\begin{figure}
\includegraphics[width=0.45\textwidth]{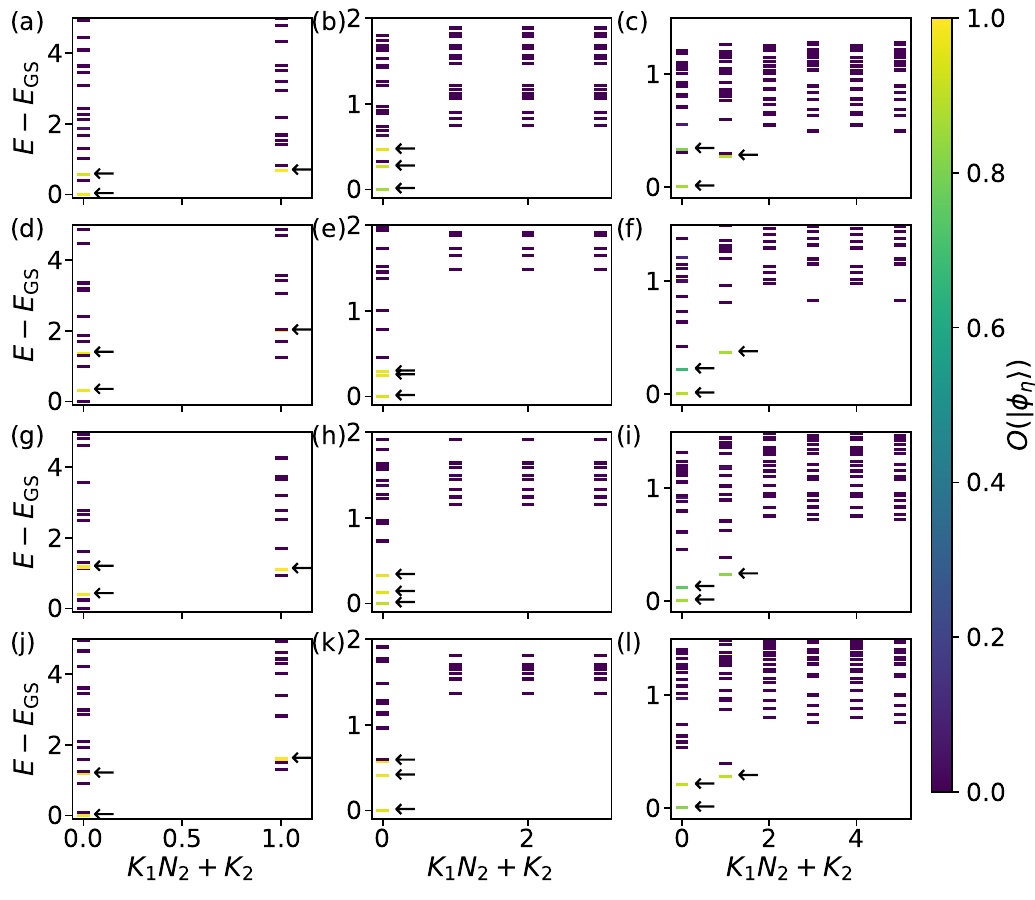}
\caption{The energy spectra for four sets of parameters from Table \ref{tab:CFTopt}: (a)-(c) global optimization with fixed $J_{\mathrm{h3}}=\mathrm{Re}J_{\mathrm{t1}}=\mathrm{Re}J_{\mathrm{t2}}=J_{\mathrm{t3}}=0$ (first row of Table \ref{tab:CFTopt} (a)),
(d)-(f) global optimization with all parameters included (last row of Table \ref{tab:CFTopt} (a)), 
(g)-(i) local optimization with all $\mathrm{Re}J_{\mathrm{t1}}=\mathrm{Re}J_{\mathrm{t2}}=J_{\mathrm{t3}}=0$ (first row of Table \ref{tab:CFTopt} (c)), and 
(j)-(l) local optimization with all parameters included (last row of Table \ref{tab:CFTopt} (c)). 
The left, middle and right columns correspond to $1\times 2$, $2\times 2$ and $3\times 2$ systems, respectively. The energy is measured with respect to the ground state energy $E_{\mathrm{GS}}$. $K_1$ and $K_2$ are the total integer lattice momenta in the directions of lattice vectors $\mathbf{a}_{1}$, $\mathbf{a}_{2}$, respectively, with $K_1\in \{0,1,\dots, N_1-1\}$ and $K_2\in \{0,1,\dots, N_2-1\}$. The color denotes the squared overlap $O(\ket{\phi_\eta})$ \eqref{eq:overlap} for the given state $\ket{\phi_\eta}$. The black arrows point to the three states with highest squared overlap in each subfigure.}
\label{fig:optspectra}
\end{figure}

From the first row of Table \ref{tab:CFTopt} (a), it can be seen that one can get good squared overlaps ($O>0.8$ for all systems and all three states $\ket{\phi_1}$, $\ket{\phi_2}$, $\ket{\phi_3}$) even when four parameters are neglected. However, in such a case, the topological degeneracy is not visible for any of the three considered system sizes, as can be seen in the energy spectra in Fig.\ \ref{fig:optspectra} (a)-(c). We do not rule out the possibility that it will appear in larger systems. 

As an aside note, we remark that the case corresponding to the first row of Table \ref{tab:CFTopt} (a) can serve as a justification for why we chose to optimize the maximum squared overlaps within the 20 lowest-energy states of the $\mathbf{K}=[0,0]$ subspace, instead of just the squared overlaps of the three lowest  states. Using the former procedure, we are able to identify the parameters for which the system seems to be at least in the proximity of a topological phase. In contrast, in the latter approach, the optimal squared overlaps of the three lowest states in the $\mathbf{K}=[0,0]$ subspace for a $2\times 2$ system are 0.477, 0, 0.766 (note that the second state has zero overlap despite the fact that all the model states have $\mathbf{K}=[0,0]$). That is, the spectrum still lacks the topological degeneracy, and the squared overlaps are worse than in the first row of Table \ref{tab:CFTopt} (a). A possible reason is that if we look only at the three lowest states, for some of them the overlaps may vanish at the datapoints considered in the optimization procedure. In consequence, in such a way we neglect some information that can direct the optimization process towards the topological phase, or at least the proximity of it.

The topological degeneracy can be seen for $2\times 2$ and $3\times 2$ systems when more parameters are included. Examples can be seen in the further subfigures of Fig.\ \ref{fig:optspectra}: Fig.\ \ref{fig:optspectra} (d)-(f), corresponding to the last row of Table \ref{tab:CFTopt} (a), i.e. a global optimization with all the parameters included; Fig.\ \ref{fig:optspectra} (g)-(i), referring to the first row of Table \ref{tab:CFTopt} (c), i.e. a local optimization with $\mathrm{Re}J_{\mathrm{t1}}=\mathrm{Re}J_{\mathrm{t2}}=J_{\mathrm{t3}}=0$; and Fig.\ \ref{fig:optspectra} (j)-(l), corresponding to the last row of Table \ref{tab:CFTopt} (c), i.e. a local optimization with all the parameters included. The fact that we do not observe the topological degeneracy in $1\times 2$ systems is not surprising, as for such small systems the finite size effects are particularly large.

Comparing Table \ref{tab:CFTopt} (a) to Table \ref{tab:CFTopt} (c), we can see that the additional local optimization sometimes increases the squared overlaps not only in the $2\times 2$ systems (which it does by definition), but also in the $3\times 2$ systems -- compare e.g. the last lines of Tabs.\ \ref{tab:CFTopt} (a) and (c) or Figs.\ \ref{fig:optspectra} (f) and (l). However, this is not always the case -- compare e.g. the penultimate rows of Tabs.\ \ref{tab:CFTopt} (a) and (c). Also, the local optimization does not necessarily increase the energy gap above the three states with high overlaps -- see the difference between Fig.\ \ref{fig:optspectra} (e), where the gap is visible, and Fig.\ \ref{fig:optspectra} (k), where it is almost nonexistent. Such behavior is not surprising, as we expect the results to be influenced by finite-size effects. Thus, it seems that whether or not the optimization reaches the actual global minimum is not important -- it is enough if the result is reasonably close to it, and lies in the region of the phase diagram corresponding to the Moore-Read spin liquid phase. Actually, a failure of the SHGO procedure can sometimes be helpful, because if a given parameter is zero while still yielding relatively good squared overlaps, as in the third and fifth rows of Table \ref{tab:CFTopt}, one may decide to neglect it and simplify the Hamiltonian (although we note that the $3\times 2$ systems corresponding to rows 3 and 5 of Table \ref{tab:CFTopt} lack topological degeneracy).

We note that there is an interesting alternative to the ``brute-force optimization'' approach used by us here. It is based on the diagonalization of the covariance matrix (or another, related matrix). This method was used in \cite{greiter2009nonabelian} and explained in more detail in \cite{chertkov2018computational,greiter2018method}. We believe that such an approach can be useful in the further studies of approximate parent Hamiltonians in our system.

\begin{table*}[htbp]
{
(a) Global\\
\begin{ruledtabular}
\begin{tabular}{lrrrrrrrrrrrrrrrrrr}
\multicolumn{1}{l}{Bnds} & \multicolumn{8}{l}{Parameters}  & \multicolumn{9}{l}{Squared overlaps} & \multicolumn{1}{l}{TD} \\ 
& \multicolumn{1}{l}{$J_{\mathrm{s1}}$} & \multicolumn{1}{l}{$J_{\mathrm{h2}}$} & \multicolumn{1}{l}{$J_{\mathrm{h3}}$} & \multicolumn{1}{l}{$\mathrm{Re}J_{\mathrm{t1}}$} & \multicolumn{1}{l}{$\mathrm{Im}J_{\mathrm{t1}}$} & \multicolumn{1}{l}{$\mathrm{Re}J_{\mathrm{t2}}$} & \multicolumn{1}{l}{$\mathrm{Im}J_{\mathrm{t2}}$} & \multicolumn{1}{l}{$J_{\mathrm{t3}}$}
& \multicolumn{3}{l}{$1\times 2$}  & \multicolumn{3}{l}{$2\times 2$} & \multicolumn{3}{l}{$3\times 2$}  &\\ \hline
B & -0.128 & 0.153 & NA & NA & 0.190 & NA & 0.147 & NA & 0.998 & 0.975 & 0.937 & 0.960 & 0.926 & 0.867 & 0.886 & 0.867 & 0.808 & $\times$~$\times$~$\times$\\
A & -0.309 & -0.510 & -0.871 & NA & 0.500 & NA & 0.629 & NA & 0.998 & 0.986 & 0.984 & 0.955 & 0.942 & 0.909 & 0.845 & 0.719 & 0.549 & $\times$~$\checkmark$~$\checkmark$\\
B & -0.500 & 0.000 & -0.295 & -0.188 & 0.391 & NA & 0.418 & NA & 0.999 & 0.972 & 0.965 & 0.965 & 0.946 & 0.902 & 0.862 & 0.835 & 0.770 &$\times$~$\checkmark$~$\times$\\
B & NA & 0.141 & NA & 0.101 & 0.147 & -0.029 & 0.140 & NA & 0.998 & 0.986 & 0.864 & 0.964 & 0.951 & 0.895 & 0.900 & 0.894 & 0.883 & $\times$~$\checkmark$~$\checkmark$\\
A & NA & 0.000 & -0.608 & 0.179 & 0.500 & -0.138 & 0.576 & NA & 0.967 & 0.896 & 0.896 & 0.919 & 0.901 & 0.857 & 0.804 & 0.789 & 0.719 & $\times$~$\checkmark$~$\times$\\
B & NA & 0.141 & NA & 0.119 & 0.153 & -0.056 & 0.155 & -0.028 & 1.000 & 0.989 & 0.826 & 0.964 & 0.952 & 0.900 & 0.905 & 0.898 & 0.880 & $\times$~$\checkmark$~$\checkmark$\\
B & -0.299 & -0.043 & -0.322 & 0.000 & 0.320 & -0.066 & 0.415 & NA & 1.000 & 0.980 & 0.936 & 0.971 & 0.966 & 0.935  & 0.906 & 0.899 & 0.852 & $\times$~$\checkmark$~$\checkmark$\\
B & -0.500 & -0.632 & -1.000 & -0.132 & 0.461 & 0.154 & 0.600 & 0.255 & 0.994 & 0.978 & 0.975 & 0.970 & 0.965 & 0.945 & 0.896 & 0.867 & 0.672 & $\times$~$\checkmark$~$\checkmark$\\ 
\end{tabular}
\end{ruledtabular}
~\\~\\
(b) Bounds for global optimization\\
\begin{ruledtabular}
\begin{tabular}{lllllllll}
& \multicolumn{1}{l}{$J_{\mathrm{s1}}$} & \multicolumn{1}{l}{$J_{\mathrm{h2}}$} & \multicolumn{1}{l}{$J_{\mathrm{h3}}$} & \multicolumn{1}{l}{$\mathrm{Re}J_{\mathrm{t1}}$} & \multicolumn{1}{l}{$\mathrm{Im}J_{\mathrm{t1}}$} & \multicolumn{1}{l}{$\mathrm{Re}J_{\mathrm{t2}}$} & \multicolumn{1}{l}{$\mathrm{Im}J_{\mathrm{t2}}$} & \multicolumn{1}{l}{$J_{\mathrm{t3}}$}\\
\hline
A & (-1,1) & (-1,1) & (-1,1) & (-1,1) & (0,1) & (-1,1) & (0,1) & (-1,1) \\
B & (-1,0) & (-1,1) & (-2,0) & (-0.6,0.6) & (0,0.6) & (-0.6,0.2) & (0,1.2) & (-0.6,0.6) \\
\end{tabular}
\end{ruledtabular}
~\\~\\
(c) Local\\ 
\begin{ruledtabular}
\begin{tabular}{rrrrrrrrrrrrrrrrrr}
 \multicolumn{8}{l}{Parameters}  & \multicolumn{9}{l}{Squared overlaps} & \multicolumn{1}{l}{TD} \\ 
 \multicolumn{1}{l}{$J_{\mathrm{s1}}$} & \multicolumn{1}{l}{$J_{\mathrm{h2}}$} & \multicolumn{1}{l}{$J_{\mathrm{h3}}$} & \multicolumn{1}{l}{$\mathrm{Re}J_{\mathrm{t1}}$} & \multicolumn{1}{l}{$\mathrm{Im}J_{\mathrm{t1}}$} & \multicolumn{1}{l}{$\mathrm{Re}J_{\mathrm{t2}}$} & \multicolumn{1}{l}{$\mathrm{Im}J_{\mathrm{t2}}$} & \multicolumn{1}{l}{$J_{\mathrm{t3}}$}
& \multicolumn{3}{l}{$1\times 2$}  & \multicolumn{3}{l}{$2\times 2$} & \multicolumn{3}{l}{$3\times 2$}  &\\ \hline
-0.210 & -0.165 & -0.387 & NA & 0.311 & NA & 0.367 & NA & 1.000 & 0.985 & 0.983 & 0.968 & 0.950 & 0.901 & 0.865 & 0.827 & 0.754 & $\times$~$\checkmark$~$\checkmark$\\
-0.496 & -0.214 & -0.530 & -0.154 & 0.443 & NA & 0.520 & NA & 1.000 & 0.979 & 0.977 & 0.966 & 0.950 & 0.906 & 0.826 & 0.823 & 0.766 & $\times$~$\checkmark$~$\checkmark$ \\
NA & -0.039 & -0.213 & 0.136 & 0.197 & -0.039 & 0.262 & NA & 1.000 & 0.986 & 0.974 & 0.971 & 0.961 & 0.925 & 0.912 & 0.896 & 0.844 & $\times$~$\checkmark$~$\checkmark$ \\
-0.806 & -0.019 & -0.462 & -0.251 & 0.519 & -0.105 & 0.655 & NA & 1.000 & 0.973 & 0.903 & 0.969 & 0.967 & 0.941 & 0.897 & 0.885 & 0.864 & $\times$~$\checkmark$~$\checkmark$ \\
-0.317 & -0.264 & -0.549 & -0.059 & 0.330 & 0.069 & 0.445 & 0.148 & 0.997 & 0.973 & 0.970 & 0.975 & 0.974 & 0.947 & 0.903 & 0.902 & 0.824 & $\times$~$\checkmark$~$\checkmark$\\
\end{tabular}
\end{ruledtabular}
}
\caption{(a) The results of the global optimization of $F(\mathbf{J})$. The column ``Bnds'' denotes which set of bounds is used. The squared overlaps shown here are $O(\ket{\phi_\eta})$, defined by Eq. \eqref{eq:overlap}, for the three states $\ket{\phi_\eta}$ with $\eta=1,2,3$ with the three largest $O(\ket{\phi_\eta})$ (note that they are sorted according to the squared overlap in decreasing order). The column ``TD'' shows the presence ($\checkmark$) or absence ($\times$) of topological degeneracy (see text for details). The first, second and third symbol in each row correspond to the $1\times 2$, $2\times 2$ and $3\times 2$ systems, respectively. (b) The two sets of bounds used in (a). (c) The results of local optimization of $F(\mathbf{J})$, with results from (a) as initial conditions. The cases not included in (c) are the ones in which the result from (a) is already a local minimum.}
\label{tab:CFTopt}
\end{table*}

\section{Discussion} \label{sec:discussion}
In Table \ref{tab:CFTopt} (a),(c) we presented the results of many optimization runs. A number of them display relatively good squared overlaps and exhibit topological degeneracy, and thus might serve as a starting point for a further numerical and experimental search for the Moore-Read spin liquid. We want to highlight two cases that might be particularly useful. The last row of Table \ref{tab:CFTopt} (c), corresponding to Fig.\ \ref{fig:optspectra}(j)-(l), contains the best results in terms of squared overlaps in $3\times 2$ systems, so this parameter set may be most likely to exhibit good overlaps in bigger systems. The second example is the first row of Table \ref{tab:CFTopt} (c), corresponding to Fig.\ \ref{fig:optspectra}(g)-(i). In this case, three parameters can be set to zero, and thus it can be easier to implement numerically and experimentally.
It also displays a robust energy gap, in the sense that the ratio of the gap between the third and the fourth state to the energy split between the first and the third state is high (compared to other optimization runs) in both $2\times 2$ and $3\times 2$ systems.

Let us compare our results to earlier works investigating lattices other than kagome. Greiter and Thomale \cite{greiter2009nonabelian} considered a triangular lattice model with ten independent real parameters (compared to eight or less parameters in our case), with interactions up to third neighbor (as in our case). For a 16-site cluster, they obtained overlaps 0.934,  0.959, 0.964. Squaring these numbers to get the squared overlaps yields 0.872, 0.919, 0.929. A naive linear interpolation between our results for a 12- and 18-site systems from the last row of Tab \ref{tab:CFTopt} (c) yields 0.865, 0.926, 0.927 for a nonexistent 16-site system, which is comparable to the results in \cite{greiter2009nonabelian}. Also, the energy gap above the third lowest state (measured relative to the energy split of the three lowest states) in an 18-site system in Fig.\ \ref{fig:optspectra} (l) (but not in the 12-site system in Fig.\ \ref{fig:optspectra} (k)) seems wider than in \cite{greiter2009nonabelian}. It can be further increased at the expense of decreasing overlaps, see Fig.\ \ref{fig:optspectra} (h),(i).

Glasser et al.\ \cite{glasser2015exact} studied a model on a square lattice, with only five real parameters and interactions up to third neighbor. In our case, five was the smallest number of parameters for which topological degeneracy was seen. In \cite{glasser2015exact}, the squared overlaps for a 16-site system on a plane and cylinder is over 0.94, while the torus was not considered.

\section{Conclusions}\label{sec:conclusions}
We used exact diagonalization to search for a non-Abelian chiral spin liquid on the kagome lattice. Employing a combination of the CFT methods and the numerical optimization, we found approximate parent Hamiltonians for a lattice Moore-Read state in finite toroidal systems. The spin liquid was identified by computing squared overlaps with the model states. 

In contrast to the $\nu=1/2$ Laughlin states \cite{nandy2019truncation}, for the $\nu=1$ Moore-Read quantum spin liquid the CFT truncation itself was not enough to construct a good parent Hamiltonian, and the numerical optimization was necessary to improve the result. For all of the shown optimized versions of the Hamiltonian, there are three low-energy states with squared overlaps higher than about $0.5$. In the best cases, the squared overlaps for all the three states exceed 0.8 for a $3\times 2$ system and 0.9 for a $2\times 2$ system. The high-overlap states are often the three lowest states for $3\times 2$ and $2\times 2$ systems, which we interpret as a sign of topological degeneracy even though the gap above these states is typically not large compared to the energy split between them. We have also shown that some parameters included in the truncated CFT Hamiltonian are not necessary for the creation of the spin liquid, and thus the final Hamiltonian can be simpler than predicted from the CFT truncation.

Our results suggest that it is possible to generate a non-Abelian chiral spin liquid in finite kagome systems. While the three investigated system sizes are not enough to determine the existence of the spin liquid phase in the thermodynamic limit, they are relevant for quantum simulators, where the number of sites can be comparable (e.g. 31 sites in Ref.\ \cite{satzinger2021realizing} and less in earlier studies \cite{lu2009demonstrating,song2018demonstration}). Our Hamiltonian can also be used as a starting point for tensor network calculations in the thermodynamic limit \cite{chen2018nonabelian}.

We also note that the term $S^{a}_jS^{a}_k S^{b}_j S^{b}_l$ with purely imaginary coefficient is equivalent to $\mathbf{S}_j \cdot(\mathbf{S}_k\times \mathbf{S}_l)$. Thus, the special case of our Hamiltonian with only $J_{\mathrm{h1}}$, $J_{\mathrm{s1}}$ and $\mathrm{Im}J_{\mathrm{t1}}$ nonzero corresponds to the Hamiltonian from Ref.\ \cite{liu2018nonabelian}. However, we found that keeping only the terms considered in \cite{liu2018nonabelian} -- nearest neighbours and smallest triangles -- was not enough to guarantee good overlaps with the model states. This suggests that adding more distant chiral terms to the Hamiltonian from Ref.\ \cite{liu2018nonabelian} may improve the chances of finding the non-Abelian chiral spin liquid.

We note that large overlaps with model states are not a necessary condition for the existence of a chiral spin liquid. The model states \eqref{eq:modelstate} are just a single example of a whole topological class of wavefunctions, thus it might be the case that the ground states have the correct topological order even when the overlaps with these states are small. Thus, it could be that our Hamiltonian can be further simplified while still generating the Moore-Read spin liquid.

\begin{acknowledgments}
We thank Didier Poilblanc, Juraj Hasik, Matthieu Mambrini and Sylvain Capponi for fruitful discussions. This work has been supported by the Independent Research Fund Denmark under grant number 8049-00074B and the Carlsberg Foundation under grant number CF20-0658.
\end{acknowledgments}

\bibliography{kagome}
\end{document}